\documentclass[twocolumn,pra,superscriptaddress,aps,10pt]{revtex4-2}
\usepackage{graphicx,subfigure}          
\usepackage{amsfonts,amsmath,amssymb,amsthm}    
\usepackage[dvipsnames]{xcolor}          
\usepackage[
colorlinks=true,
linkcolor=blue,
citecolor=blue,
urlcolor=blue]{hyperref}                
\usepackage{comment}                    
\usepackage{appendix}
\usepackage{enumerate}
\usepackage{booktabs}

\newtheorem{lemma}{Lemma}
\newtheorem{definition}{Definition}
\newtheorem{remark}{Remark}
\newtheorem{theorem}{Theorem}

\definecolor{fg}{rgb}{0.13, 0.55, 0.13}
\newcommand{\argp}[1]{\left( #1 \right)}
\newcommand{\args}[1]{\left[ #1 \right]}
\newcommand{\argc}[1]{\left\{ #1 \right\}}
\newcommand{\ket}[1]{\left\vert #1 \right\rangle}
\newcommand{\bra}[1]{\left\langle #1\right\vert}

\newcommand{\abs}[1]{\left\vert #1 \right\vert}
\newcommand{\norm}[1]{\left\Vert #1 \right\Vert_p}

\newcommand{\NC}{\mathcal{N}^{p,F_L}_{\mathcal{C}}}
\newcommand{\NCs}{\mathcal{N}^{p,W_s}_{\mathcal{C}_g}}
\newcommand{\NCwo}{\mathcal{N}^{1,W}_{\mathcal{C}_g}}

\newcommand{\Tr}{\mathrm{Tr}}
\newcommand{\dd}{\mathrm{d}}
\newcommand{\dg}{\dagger}
\newcommand{\Ccal}{\mathcal{C}}
\newcommand{\PhiL}{\Phi_{\mathrm{L}}}
\newcommand{\Pcl}{P_{\mathrm{cl}}}
\newcommand{\normp}[1]{\left\Vert #1 \right\Vert_p}

\definecolor{dgreen}{rgb}{0.0, 0.42, 0.24}

\begin{document}

\title{
No-Go Theorem for Norm-Based Nonclassicality Certification with Linear Functionals}

\author{Soumyakanti Bose}
\email{soumyakanti.bose@gmail.com}
\affiliation{NextQuantum Innovation Research Center, Department of Physics \& Astronomy, Seoul National University, Seoul 08826, Republic of Korea}
\affiliation{Department of Physics, School of Engineering and Sciences (SEAS), SRM University-AP, Amaravati 522240, Andhra Pradesh, India}

\author{Yong Siah Teo}
\email{ys\_teo@sejong.ac.kr}
\affiliation{NextQuantum Innovation Research Center, Department of Physics \& Astronomy, Seoul National University, Seoul 08826, Republic of Korea}
\affiliation{Department of Quantum Information Science and Engineering, Sejong University, Seoul 05006, Republic of Korea}

\author{Hyukjoon Kwon}
\email{hjkwon@kias.re.kr}
\affiliation{School of Computational Sciences, Korea Institute for Advanced Study, Seoul 02455, Republic of Korea}

\author{Hyunseok Jeong}
\email{jeongh@snu.ac.kr}
\affiliation{NextQuantum Innovation Research Center, Department of Physics \& Astronomy, Seoul National University, Seoul 08826, Republic of Korea}

\date{\today}

\begin{abstract}
Despite various approaches to encode nonclassical light, most existing formulations either suffer from limited practicality or rely on optimization procedures that are computationally demanding. 
Here, we develop a framework for quantifying optical nonclassicality without invoking any optimization, using generic classicalization channels, $p$-norms and linear functionals generated by displaced rotationally-invariant operators. 
However, for a subset of such measures, involving $s$-ordered distributions and a Gaussian classicalization channel, we analytically establish a no-go theorem demonstrating that no such universal measure can exist. 
We further substantiate our theoretical result through explicit examples involving both Gaussian and non-Gaussian states, followed by a generic criterion for a no-go theorem, whose the existence remains to be established.
\end{abstract}

\maketitle

\section{Introduction}
\label{sec:intro}

Optical states that cannot be represented by a bona fide positive-definite distribution over the set of classical pure states—namely, the coherent states $\ket{\alpha}$ \cite{Hillery1985}—are said to be quantum. 
In other words, any optical state that admits a decomposition $\rho = \int\frac{d^2\alpha}{\pi} P(\alpha) \ket\alpha\bra\alpha$ is classical if $P(\alpha) \ge 0$, and quantum otherwise, i.e., when $P(\alpha)$ becomes negative or more singular than a delta function \cite{Glauber1963, Sudarshan1963, Franson2018}.
Quantum optical states characterized by such nonregular sampling distributions—formally known as the Sudarshan–Glauber diagonal $P$-function \cite{Glauber1963, Sudarshan1963}—play a crucial role across modern quantum information science and technology \cite{Braunstein2005, Andersen2010, Adesso2014}, from metrology \cite{Tan2019} and teleportation \cite{Furusawa1998} to quantum communication \cite{Grosshans2002, Fabian2018, Zhang2024} and computation \cite{Llyod1999, Bartlett2003, Mari2012, Veitch2013}.

The operational interpretation of the $P$-function and its Gaussian-convolved counterparts—such as the Wigner function \cite{Wigner1920} and Husimi–Kano $Q$-function \cite{Husimi1940, Kano1965}—in terms of measurement statistics \cite{Wodkiewicz1984, Leonhardt1995, Buzek1995, Buzek1995a, Manko2000} has led to several foundational insights in quantum theory. 
These include the necessity of negativity for quantum behavior \cite{Ferrie2010} and the impossibility of classical simulation of quantum systems \cite{Pashayan2015, Semenov2021}.
Consequently, developing an operational resource theory \cite{Theurer2017, Takagi2019, Chitambar2019} of nonclassicality—encoded in the negativity or singularity of the $P$-function ($P(\alpha) \ngeq 0$)—has remained a central pursuit in recent decades \cite{Yadin2018, Kwon2019, Lami2021, Regula2021}.

Since the set of classical states forms a convex subset of the full state space, the Hahn–Banach theorem ensures the existence of a hyperplane that separates classical from quantum states. 
This naturally leads to distance-based quantifiers of nonclassicality, defined via the minimal distance from the set of classical states. Such measures have been formulated using $p$-norms \cite{Hillery1987, Dodonov2000, Dodonov2003, Nair2017, Marian2020, Park2021} and the Bures metric \cite{Marian2002}, within the Hilbert-space framework.

Analogous to quantum coherence in finite-dimensional systems \cite{Baumgratz2014, Winter2016}, researchers have also proposed resource-theoretic formulations of phase-space coherence \cite{Tan2017}. 
Combined with phase-space filtering techniques \cite{Kiesel2010, Kuhn2018}, this approach enables an operational quantification of the negativity of the Glauber–Sudarshan $P$-function \cite{Tan2020}. 
Moreover, leveraging the positive semi-definiteness of the Husimi–Kano $Q$-function \cite{Husimi1940, Kano1965}, entropic formulations of nonclassicality have been developed \cite{Bose2019}, alongside methods for identifying extremal quantum states \cite{Goldberg2020}.

Despite their mathematical elegance and conceptual depth, most existing measures either require optimization over the classical set \cite{Hillery1987, Marian2002, Park2021, Bose2019} or lack practical feasibility \cite{Tan2017, Tan2020}. 
Several such measures also fail to capture nonclassicality in general \cite{Bose2017}, precisely because they omit this optimization. 
This raises a fundamental question: Is it possible to construct a universal, optimization-free, resource-theoretic measure of nonclassicality?

In this article, we explore this possibility by formulating a distance-based, $p$-norm–based resource-theoretic measure of nonclassicality, without optimization over the set of classical states. 
To define the measure, we consider the class of linear functionals $F_L$ for which generators are displaced rotationally-invariant operators, i.e. $F_L:=\text{Tr}[\rho G_\eta]$.
We encode the nonclassicality in terms of the $p$-norm distance, $\NC$, between the linear functional ($F_L$) acting on a quantum state and its classical counterpart---obtained as the output of a generic nonclassicality-breaking/classicalization channel $\mathcal{C}$. 
While our framework serves as a bona fide nonclassicality certifier, it also generalizes existing norm-based measures within a unified formalism.

However, we further demonstrate that for a subclass of such measures, $\NCs$, defined in terms of $L^p$-norms, $s$-ordered distributions and a universal Gaussian classicalization channel $\mathcal{C}_g$, universality cannot be guaranteed. 
By considering a manifestly nonclassical state for which the $P$-function becomes negative, constructed by subtracting vacuum from a thermal state, we show that the $\NCs$ fail to detect the nonclassicality. 
This leads to a no-go theorem, which states that no universal, optimization-free, norm-based nonclassicality certifier can be constructed with $s$-ordered distributions and $\mathcal{C}_g$. 
We illustrate this theorem further using both Gaussian and non-Gaussian examples, within an $L^1$-norm framework and employing the Wigner function.

We further provide a generalized criterion for all such $p$-norm based measures, $\NC$, that respects the no-go condition for any vacuum-subtracted classical state---a much larger set of just thermal states. 
However, unlike the $\NCs$, the \emph{proof of existence} for the no-go condition for $\NC$ remains open.

\section{Classicalization channel and Nonclassicality Measure}
\label{sec:cchannel_measure}

\begin{definition} 
    \label{def:classicalchannel}
    A \emph{classicalization}-channel $\mathcal{C}$ is defined as a map that transforms any given state into a classical state such that, for any arbitrary density operator, $\rho = \int\frac{d^2\alpha}{\pi} P(\alpha) \ket\alpha\bra\alpha$,
    \begin{align}
        \mathcal{C}(\rho) &= \int\frac{d^2\beta}{\pi} P_\text{cl}(\beta) \ket\beta\bra\beta,
        \label{eq:action_class_channel}
    \end{align}
    where $P_\text{cl}(\beta) = \int\frac{d^2\alpha}{\pi} P(\alpha) f(\beta - \alpha) \geq 0$ such that $f(\beta - \alpha)$ is the smoothing function applied on $P(\alpha)$.
\end{definition}
The existence of such mapping may be appreciated in terms of the well-known results such as {\bf (a)} a Gaussian smoothing---with the ground state Wigner function---that always yields a positive regular $Q$ distribution irrespective of the $P$ function and {\bf (b)} any Wigner negative state becomes Wigner positive at $50\%$ photon-loss \cite{Cahill1969}.

\textbf{Norm-based measure of nonclassicality without optimization}---As a bona fide distance measure we consider the $p$-norm between the linear functional ($F_L$) of $\rho$ and $\mathcal{C}(\rho)$, such that $F_L\argp{\sum_k p_k \rho_k} = \sum_k p_k F_L(\rho_k)$.
The image of the functional $F_L$ could be both operators and functions, depending on whether it maps the density operator onto Hilbert space or phase-space.
Consequently, we define the measure as
\begin{equation}
    \NC(\rho) = \left\Vert F_L(\rho) - F_L[\mathcal{C}(\rho)] \right\Vert_p,
    \label{eq:nrho_def}
\end{equation}
where $\Vert A \Vert_p = \argc{ \text{Tr}\args{ \argp{\sqrt{A^\dagger A}}^p} }^{1/p}$.
It may be readily appreciated that, as a consequence of $p$-norm, $\mathcal{N}_\mathcal{C}(\rho)$ satisfies the positivity, symmetry and triangle inequality which are necessary for a bona fide distance measure (Appendix \ref{append_sec:props}).

Here, we consider a class of linear functionals which are generated from a displaced rotationally-invariant operators, i.e., $F_L:=F_L^{G(\alpha)} = \text{Tr} \args{\rho\,G(\alpha)}$, where $G(\alpha) = D(\eta)G_0D^\dagger(\eta)$ such that $\args{G_0,R(\theta)} = 0$, $R(\theta)$ being the phase-space rotation operator.
Such a choice stems from {\bf (a)} all phase-space distributions fall under this category and {\bf (b)} any bona-fide nonclassicality measure must be independent of classical operations such as phase-space rotation and displacement (Appendix \ref{append_sec:invariance}).
We now prove the rest of the properties as follows.

\subsection{Resource theoretic properties of $\NC$}

\textbf{Convexity}---Let us consider a convex mixture of state given by $\lbrace p_k,\rho_k \rbrace$, such that $0\leq p_k\leq 1$ and $\sum_k p_k = 1$ for which $\mathcal{C}(\rho) = \sum_k p_k \mathcal{C}(\rho_k)$.
Then by using the convexity of $\mathcal{L}^p$ norm ($p\geq 1$) we get
\begin{align}
    \mathcal{N}_\mathcal{C}^{p, F_L}\argp{ \rho } &= \left\Vert F_L\argp{\sum_k p_k \rho_k} - F_L\args{\sum_k p_k \mathcal{C}(\rho_k)} \right\Vert_p
    \nonumber 
    \\
    &\leq \sum_k p_k \left\Vert F_L(\rho_k) - F_L\args{\mathcal{C}(\rho_k)} \right\Vert_p
    \nonumber 
    \\
    &= \sum_k p_k \mathcal{N}_\mathcal{C}^{p, F_L}(\rho_k).
    \label{eq:nrho_convexity}
\end{align}

\textbf{Strong monotonicity}---
Let us define a \emph{free selective operation as an instrument} $\mathcal{E}_k$ such that
\begin{equation}
    \mathcal{E}_k(\rho) = \int d\mu P_\text{cl}(\mu) q_k(\mu)\,V(\mu) \rho V^\dagger(\mu),
    \label{eq:instrument}
\end{equation}
where $0\leq q_k(\mu)\leq 1$ is the conditional probability of announcing outcome $k$ given the realized parameter $\mu$ such that  $\sum_k q_k(\mu)= 1 ~\forall\mu $.
$V(\mu) = R(\theta)D(\lambda)$ are the rotation--displacement unitary operators. 
Equivalently, the Kraus operators are
\begin{equation}
    K_{k,\mu}=\sqrt{P_\text{cl}(\mu) q_k(\mu)} V(\mu), \quad \sum_k \int d\mu K_{k,\mu}^\dagger K_{k,\mu}=\mathbb{I},
    \label{eq:kraus}
\end{equation}
\emph{i.e.}, each Kraus operator is proportional to a free unitary. 

As a consequence, in a straightforward calculation one can show that (Appendix \ref{append_sec:strong})
\begin{align}
        &\sum_k p_k \mathcal{N}_\mathcal{C}^{p, F_L}(\rho_k) = \sum_k p_k \left\Vert
        F_L\argp{\rho_k} -F_L\args{\mathcal{C}(\rho_k)} 
        \right\Vert_p 
        \nonumber 
        \\
        &\leq \sum_k \int d\mu P_\text{cl}(\mu) q_k \left\Vert 
        F_L\argp{ \rho } - F_L\args{\mathcal{C}\argp{ \rho } } 
        \right\Vert_p
        \nonumber 
        \\
        &= \int d\mu P_\text{cl}(\mu) \argp{ \sum_k q_k} \mathcal{N}_\mathcal{C}^{p, F_L}(\rho)
        = \mathcal{N}_\mathcal{C}^{p, F_L}(\rho).
        \label{eq:nrho_strongmonotonicity}
\end{align}

It may further be noted that \emph{the current model of free selective operations cannot be extended to the ancilla-assisted linear optics, followed by projective measurement} (Appendix \ref{append_sec:strong}).

\textbf{Weak monotonicity}---
For a non-selective free operation, $\Phi_L = \sum_k\mathcal{E}_k$, convexity and strong-monotonicity together imply
\begin{align}
    \NC[\Phi_L(\rho)] &= \NC\argp{\sum_k p_k \rho_k}
    \nonumber 
    \\
    &\leq \sum_k p_k \NC(\rho_k)
    \leq \NC(\rho).
    \label{eq:nrho_weakmonotonicity}
\end{align}

Nonetheless, the invariance of $\mathcal{N}_\mathcal{C}^{p, F_L}$ under rotation and displacement operations (Appendix \ref{append_sec:invariance}) yields $\mathcal{N}_\mathcal{C}^{p, F_L} \argp{\ket\alpha} = \mathcal{N}_\mathcal{C}^{p, F_L} \argp{\ket 0}$. 
This, together with Young's convolution inequality, leads to for a test state, $\rho_\text{test}=\int\frac{d^2\alpha}{\pi} P_\text{test}(\alpha) \ket\alpha\bra\alpha$,
\begin{align}
    \mathcal{N}_\mathcal{C}^{p, F_L}\argp{\rho_\text{cl}} &\leq \left\Vert P_\text{test}(\alpha) \right\Vert_1 ~\mathcal{N}_\mathcal{C}^{p, F_L}\argp{\ket 0},
    \label{eq:nrho_classmaxval}
\end{align}
implying $\mathcal{N}_\mathcal{C}^{p, F_L}\argp{\rho_\text{cl}} \leq \mathcal{N}_\mathcal{C}^{p, F_L}(\ket 0), ~\forall P_{\text{cl}} \geq 0$.
In other words, $\mathcal{N}_\mathcal{C}^{p, F_L}\argp{\rho_\text{cl}} > \mathcal{N}_\mathcal{C}^{p, F_L}(\ket 0)$ holds if and only if $\lVert P_{\text{test}}(\alpha) \rVert_1 > 1$, that is, when $P_{\text{test}}(\alpha)$ either takes negative values or becomes a singular function for which the norm diverges.

As a direct consequence, we define a convex, resource-theoretic measure of nonclassicality as 
\begin{equation}
    \mathcal{M}(\rho) = \mathcal{N}_\mathcal{C}^{p, F_L}(\rho) - \mathcal{N}_\mathcal{C}^{p, F_L}(\ket 0),
    \label{eq:measure}
\end{equation}
where $\mathcal{M}(\rho)\leq 0$ for all classical states and $\mathcal{M}(\rho) > 0$ signifies nonclassicality such that $\mathcal{M}(\rho)=0$ only for the classical pure states, i.e., the coherent states.

In analogy with the $L^p$-norm distance between two functions $f(x)$ and $g(x)$, $\mathcal{D}_{f,g}^p = \left\Vert f(x) - g(x) \right\Vert_p = \args{ \int_{\Omega} d\mu(x) \abs{f(x) - g(x)}^p }^{1/p}$, which quantifies their statistical distinguishability, one can interpret the quantity $\mathcal{N}_{\mathcal{C}}^{p,F_L}(\rho)$ as the distinguishability between a quantum state $\rho$ and its classical counterpart $\mathcal{C}(\rho)$, obtained as the output of a nonclassicality-breaking channel.
Consequently, a positive value of $\mathcal{M}(\rho)$ directly quantifies the extent to which the state $\rho$ differs from one that can be statistically represented as a mixture of coherent states.
Equivalently, $\mathcal{M}(\rho) > 0$ admits an operational interpretation as a measure of the quantum superposition—or the degree of nonclassicality—present in $\rho$.

\section{Gaussian classicalization channel and No-go theorem}
\label{sec:nogo}

\begin{definition}
	\label{def:gaussian_classical}
	A Gaussian noise-additive classicalization channel, $\mathcal C_g$, is a subset of all classicalization channels $\mathcal{C}$ \eqref{eq:action_class_channel}, defined as the quantum-limited attenuator--amplifier composite channel, i.e., $\mathcal C_g=A_2\circ E_{1/2}$, where
    \begin{align}
    \mathcal{E}_\lambda \argp{\rho} &= \rm{Tr}_\text{anc} \args{ \mathsf{B}(\lambda) \argp{ \rho\otimes \ket{0_\text{anc}}\bra{0_\text{anc}} } \mathsf{B}^{\dagger}(\lambda) } ~~\rm{and}
    \nonumber 
    \\
    \mathcal{A}_k \argp{\rho} &= \rm{Tr}_\text{anc} \args{ \mathsf{S}(\mathit{k}) \argp{ \rho\otimes \ket{0_\text{anc}}\bra{0_\text{anc}} } \mathsf{S}^{\dagger}(\mathit{k}) },
    \label{eq:defn_attenuate_amplify}
\end{align}
such that $\mathsf{B}(\lambda) = e^{\cos^{-1}\sqrt{\lambda},(a^\dagger b - a b^\dagger)}$ is a passive beam-splitter with transmittivity $\lambda$, $\mathsf{S}(k) = e^{\cosh^{-1}\sqrt{k}(a^\dagger b^\dagger - a b)}$ is a two-mode correlated squeezing operator, and $\ket{0_\text{anc}}$ denotes the vacuum state in the ancilla mode.
\end{definition}
For $\mathcal{C}_g$, $f(\beta)=e^{-|\beta|^2}$ such that $P_{\mathcal{C}_g(\rho)}(\alpha) = Q_\rho(\alpha)$, the Husimi function of $\rho$ --- equivalently, $P$ convolved with a Gaussian of variance~$1$ \cite{Linowski2024}.
This makes $\mathcal{C}_g$ the universal Gaussian classicalization channel. 

\begin{theorem}
    For the $s$-ordered quasidistribution $W_s(\rho;\alpha)=\mathrm{Tr}[\rho\,\mathcal{K}_s(\alpha)]$, the $L^p$ measure of nonclassicality $\NCs(\rho):=\norm{W_s(\rho;\alpha) - W_s[\mathcal{C}_g(\rho);\alpha]}$ cannot hold universally: there exists a quantum state $\rho_{\rm neg}^{\rm test}$ that is genuinely $P$-negative yet satisfies $\NCs(\rho) < \NCs(|0\rangle)$.
\end{theorem}

\begin{proof}---Let us consider a vacuum subtracted thermal state as
	\begin{equation}
		\rho_{\rm neg}^{\rm test} := \left(1+\frac1{\bar n}\right)\rho_{th}(\bar n) - \frac1{\bar n}|0\rangle\langle0|\,
        \label{eq:rhoneg}
	\end{equation}
	that represents a normalized physical state for every $\bar n>0$, with
	\begin{equation}
		P_{\rho_{\rm neg}^{\rm test}}(\alpha) = \frac{1+\bar n}{\bar n^2}\,e^{-|\alpha|^2/\bar n} - \frac{\pi}{\bar n}\,\delta^{(2)}(\alpha)\,,
        \label{eq:Pneg}
	\end{equation}
	carrying a negative delta spike at the origin for every finite $\bar n>0$: $\rho_{\rm neg}^{\rm test}$ is genuinely
	$P$-negative, for every $\bar n$, independent of $s$.

    It can be shown that $\forall ~s \in [-1,1)$, $\forall ~1\leq p< \infty$, $\exists ~\bar{n}^{*}(p,s)$ such that (Appendix \ref{append_sec:pneg})
    \begin{equation}
        \NCs\argp{\rho_\text{th}(\bar{n})} < \frac{\bar{n}-1}{\bar{n}+1} \NCs(\ket 0), \quad \forall ~\bar{n}>n^{*}(p,s).
        \label{eq:nogocond_therm}
    \end{equation}

    As a consequence, by using the triangle inequality, $\norm{A-B}\leq \norm{A}+\norm{B}$, one can easily show (see \cite{supple})
    \begin{align}
        \NCs\argp{\rho_{\rm neg}^{\rm test}} &\leq \argp{ 1+\frac1{\bar n} } \NCs\argp{\rho_{\rm th}(\bar{n})} + \frac{1}{\bar{n}} \NCs(\ket 0)
        \nonumber 
        \\
        &< \NCs(\ket 0).
        \label{eq:nogo_rhoneg}
    \end{align}
\end{proof}
This, in turn, indicates that $\NCs$---the $L^p$ measure defined with any $s$-ordered distributions ($W_s$) and the Gaussian classicalization channel ($\mathcal{C}_g$) is not a faithful measure.

\subsection{Practical example of no-go theorem: Gaussian and non-Gaussian cases}

We further illustrate the implications of the no-go theorem \eqref{eq:nogo_rhoneg}, by considering $p=1$ and $s=0$, i.e., with $L^1$ norm and the Wigner function. 
Here, we consider two representative physical examples---one Gaussian and one non-Gaussian.

\textbf{Gaussian case}---
Let us consider a squeezed thermal state given as $\rho_\text{st} = S(r)\rho_\text{th}(\bar{n})S^\dagger(r)$, where $S(r) = e^{\frac{r}{2}\argp{a^{\dagger 2} - a^2}}$, $\rho_\text{th}(\bar{n}) = \frac{1}{1+\bar{n}}\sum_k \argp{\frac{\bar{n}}{1+\bar{n}}}^k \ket k\bra k$; $r$ being the squeezing strength and $\bar{n}$ representing the number average thermal photon.
The onset of nonclassicality in $\rho_\text{st}$ is marked whenever $r>\frac{1}{2}\ln(2\bar{n}+1)$ for which variance of the $x$-quadrature falls below $1/2$ or alternatively the $P$-function diverges.

\begin{figure}
    \centering
    \includegraphics[width=\columnwidth]{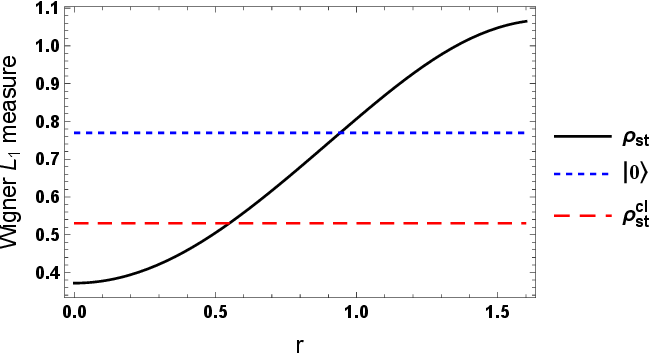}
    \caption{$\mathcal{N}_{\mathcal{C}_g}^{1,W}(\rho_\text{st})$ as a function of $r$ with $\bar{n}=1$.
    Various curves are mentioned in the legends.
    Here, $\rho_\text{st}^\text{cl}$ denotes the classicality bound for $\rho_\text{st}$, i.e., $\NCwo(\rho_\text{st})$ at $r = 0.5499$ and $\bar{n}=1$.}
    \label{fig:nrho_sqth}
\end{figure}

In Fig. \ref{fig:nrho_sqth} we plot the quantifier $\mathcal{N}_{\mathcal{C}_g}^{1,W}(\rho_\text{st})$ with the squeezing strength $r$ with average photon number set to unity, i.e., $\bar{n}=1$.
As it is evident, $\rho_\text{st}$ becomes nonclassical at $r = \frac{1}{2}\ln 3 \sim 0.5499$; however, $\mathcal{N}_{\mathcal{C}_g}^{1,W}(\rho_\text{st})$ becomes greater than $\mathcal{N}_{\mathcal{C}_g}^{1,W}(\ket 0)$ at a much higher value of $r$ ($\sim 0.95$).
This implies the state, despite being quantum, fails to manifest its nonclassicality through $\mathcal{N}_{\mathcal{C}_g}^{1,W}$ for a considerable range of squeezing strength $0.55 \leq r \lesssim 0.95$.

\begin{figure}
    \centering
    \includegraphics[width=\columnwidth]{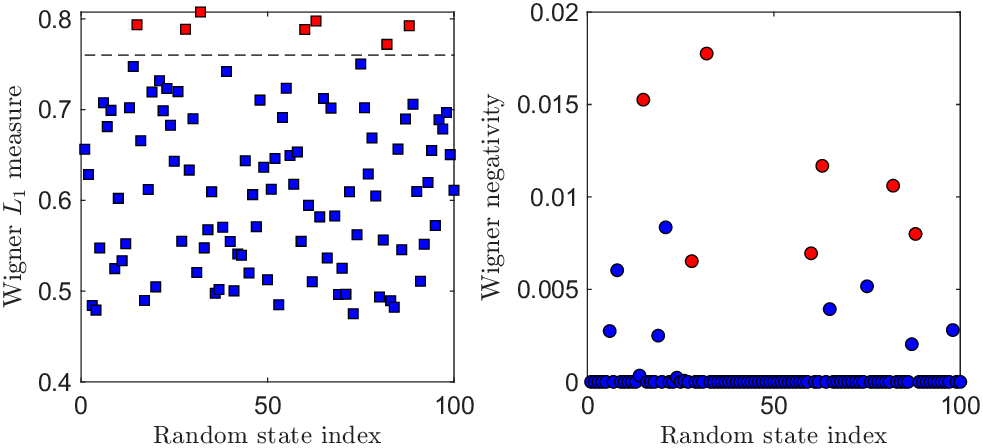}
    \caption{$\mathcal{N}_{\mathcal{C}_g}^{1,W}$ and Wigner negativity for $\rho_\text{mix}$ for $100$ different triplets $(p_0,p_1,p_2)$.}
    \label{fig:nrho_wigneg_mix}
\end{figure}

\textbf{Non-Gaussian case}---
Let us consider random mixtures of photon number states $\ket 0$, $\ket 1$ and $\ket 2$ as $\rho_\text{mix} = p_0 \ket 0\bra 0 + p_1 \ket 1 \bra 1 + p_2 \ket 2 \bra 2$ such that $p_0+p_1+p_2=1$. 
As an independent indicator of nonclassicality, we employ the Wigner negativity~\cite{Kenfack2004}, defined as $\text{Wig. neg.} = \int\frac{dx dp}{\pi} \abs{W_{\rho_\text{mix}}(x,p)} - 1$, where $W_{\rho_\text{mix}}(x,p)$ is the Wigner function for $\rho_\text{mix}$.
Figure~\ref{fig:nrho_wigneg_mix} presents results for $100$ randomly generated mixtures, comparing the nonclassicality indicated by $\mathcal{N}_{\mathcal{C}_g}^{1,W}$ with that obtained from the Wigner negativity.

In the plot of $\NCwo(\rho_\text{mix})$, the horizontal dashed line denotes the value corresponding to a coherent state.
This serves as a clear boundary separating classical and quantum cases, indicated respectively by blue and red square markers.
In the corresponding Wigner negativity plot, the states identified as quantum by $\NCwo$ are shown as red circles.
Interestingly, for a few specific combinations of $p_0,p_1$ and $p_2$, we observe nonzero Wigner negativity even though these states are classified as classical according to $\NCwo$ (blue circles in Fig.~\ref{fig:nrho_wigneg_mix}).
The specific probabilities of the failure cases are given as
\begin{equation}
	\begin{array}{ccc}
		\toprule
		p_0 & p_1 & p_2\\
		\midrule
		0.1918 & 0.3226 & 0.4857\\
		0.2686 & 0.2223 & 0.5091\\
		0.1975 & 0.3011 & 0.5014\\
		0.2611 & 0.2322 & 0.5067\\
		0.2254 & 0.2706 & 0.5040\\
		0.1699 & 0.4026 & 0.4274\\
		0.2605 & 0.2298 & 0.5097\\
		\bottomrule
	\end{array}
	\label{append_eq:table}
\end{equation}

Evidently, these cases constitute explicit examples of the \emph{no-go} behavior for non-Gaussian states---showing that certain Wigner negative states may still remain “classicalized” under the action of the $\mathcal{C}_g$.

The apparent no-go cases, $\NCwo(\rho) < \NCwo(\ket 0)$, for both the $\rho_\text{st}$ and $\rho_\text{mix}$ can further be appreciated from the operational interpretation of $\mathcal{C}_g$ which acts analogously to the quantum depth measure \cite{Lee1991}---minimum Gaussian smoothing required to map a nonregular $P$-function into a positive semi-definite distribution.

For Gaussian states, a milder channel $\mathcal{C}' = \sqrt{\mathcal{C}_g}$ \cite{Linowski2024} already converts the Glauber–Sudarshan $P$-function into the Wigner function, which is itself positive semi-definite.
Likewise, for the non-Gaussian pure states, while $\mathcal{C}_g$ represents precisely the minimum required Gaussian smoothing \cite{Lutkenhaus1995}, mixed non-Gaussian states typically require less smoothing than that implemented by $\mathcal{C}_g$.

In other words, $\mathcal{C}_g$ effectively over-smoothens both Gaussian and certain non-Gaussian mixed states and thereby suppressing their detectable quantum signatures, yields underestimation of $\NCwo(\rho)$.

\section{Conclusion}
\label{sec:conclude}

In this work, we have established a \emph{no-go} result concerning the construction of vector-norm-based measures of nonclassicality without invoking any optimization procedures.
To this end, we first developed a bona fide convex resource-theoretic measure as $\NC$, defined via the $p$-norm distance between a class of linear functionals, $F_L$, of the quantum state and the corresponding classical state obtained as the output of a nonclassicality-breaking channel, $\mathcal{C}$.
Such class of linear functionals are defined as $F_L:=\rm{Tr}[\rho G_\eta]$ where the generators $G_\eta$ are displaced rotation-invariant operators.
One may choose such linear functionals like the density operator itself or say the Husimi-function and other $p$ values such as $p=2$, which leads to the Hilbert-Schmidt distance.
While inherently encoding the coherent superposition content of the state, our formulation provides a general operational framework for norm-based nonclassicality certification whenever $\NC(\rho) > \NC(\ket 0)$.

We subsequently established the central premise of the \emph{no-go theorem}---any such measure, $\NCs$, using $L^p$-norm measure, s-ordered distribution ($W_s$) and a universal Gaussian classicalization channel ($C_g$), fails to detect the nonclassicality faithfully.
Through explicit mathematical construction of vacuum-subtracted thermal state, we have proved the existence for $P$-negative states, $\rho_\text{neg}$, for which $\NCs(\rho_\text{neg})<\NCs(\ket 0)$ despite $\rho_\text{neg}$ being manifestly nonclassical. 
This establishes the impossibility of constructing a universal/faithful norm-based nonclassicality certifier solely from s-ordered distributions.
We further illustrated this result using the $L^1$-norm, $\mathcal{C}_g$ and the Wigner functions for both Gaussian and non-Gaussian states.

While the \emph{no-go theorem} could be analytically established for arbitrary $L^p$ norm ($p\geq 1$), $s$-ordered distributions and the Gaussian channel $\mathcal{C}_g$, the validity of this result with arbitrary classicalization channel $\mathcal{C}$ and general linear functional $F_L:=\text{Tr}\argp{\rho G_\eta}$ remains an open question.
Under the action of $\mathcal{C}$, a $P$-negative state can similarly be constructed from any classical state ($\rho_\text{cl}$) as
\begin{equation}
    \rho_\text{neg}^\text{cl} = \frac{1}{1-p_0}\rho_\text{cl} - \frac{p_0}{1-p_0}\ket 0\bra 0,
\end{equation}
where $p_0 = \text{Tr}\args{\rho_\text{cl} \ket 0\bra 0}$. 
For this state the nonclassicality measure $\NC$ \eqref{eq:nrho_def} fails to be faithful whenever (see \cite{supple})
\begin{equation}
    \NC(\rho_\text{cl}) < (1-2p_0) \NC(\ket 0).
\end{equation}

However, the existence of such states satisfying this criterion remains to be established. 

\section{Acknowledgments}
\label{sec:credit}

This work was supported by the National Research Foundation of Korea (NRF) grant funded by the Korea government (MSIT) (Nos. RS-2023-00237959, RS-2024-00413957, RS-2024-00438415, RS-2023-NR076733, RS-2025-25464492, RS-2026-25590066, and RS-2026-25609336), the Institute of Information \& Communications Technology Planning \& Evaluation (IITP) grant funded by the Korea government (MSIT) (IITP-2025-RS-2020-II201606, IITP-2025-RS-2024-00437191, and RS-2025-02219034), and the Institute of Applied Physics at Seoul National University. 
H.K. is supported by the KIAS Individual Grant No. CG085302 at Korea Institute for Advanced Study. 
Y.S.T. is supported by the faculty research fund of Sejong University in 2026.



\appendix
\section{Bona fide resource-theoretic distance measure and quantification of the channel $\Ccal$}
\label{append_sec:props}

We now quantify the action of the ``classicalization'' channel $\Ccal$ on any quantum state $\rho$ in terms of a bona fide distance function between the quantum states $\rho$ and $\Ccal(\rho)$, \emph{i.e.}, $D[\rho,\Ccal(\rho)]$. 
To enforce a resource-theoretic evaluation, $D$ needs to satisfy the following properties:
\begin{itemize}
    \item[(i)] \emph{Positive semi-definiteness}: $D(\rho_1,\rho_2)\geq 0$ for any two quantum states $\rho_1$ and $\rho_2$.
    
    \item[(ii)] \emph{Symmetry}: $D(\rho_1,\rho_2)=D(\rho_2,\rho_1)$ for any two quantum states $\rho_1$ and $\rho_2$.
    
    \item[(iii)] \emph{Triangle inequality}: $D(\rho_1,\rho_2)+D(\rho_2,\rho_3)\geq D(\rho_1,\rho_3)$ for any bona fide quantum states $\rho_1$, $\rho_2$ and $\rho_3$.
    
    \item[(iv)] \emph{Convexity}: For any $\rho=\sum_k p_k\rho_k$ such that $p_k\geq0$ and $\sum_k p_k=1$, $D[\rho,\Ccal(\rho)]\leq\sum_k p_k D[\rho_k,\Ccal(\rho_k)]$.
    
    \item[(v)] \emph{Weak monotonicity}: For any linear-optical map $\PhiL$, $D\{\PhiL(\rho),\Ccal[\PhiL(\rho)]\}\leq D[\rho,\Ccal(\rho)]$.
    
    \item[(vi)] \emph{Strong monotonicity}: For any \emph{free selective operation} $\{\mathcal{E}_k\}$, i.e., an instrument whose Kraus operators are each proportional to a rotation--displacement unitary, $K_{k,\mu}=\sqrt{\Pcl(\mu)\,q_k(\mu)}\,V(\mu)$ with $q_k(\mu)\geq0$ and $\sum_k q_k(\mu)=1$ (see Definition~\ref{def:free} in Sec.~\ref{append_sec:strong}) --- one has
    \begin{equation}
        \sum_k p_k\, D[\rho_k,\Ccal(\rho_k)]\;\leq\; D[\rho,\Ccal(\rho)]\,,
    \end{equation}
    where $\rho_k=\mathcal{E}_k(\rho)/p_k$ and $p_k=\Tr[\mathcal{E}_k(\rho)]$.
\end{itemize}
As shown by the explicit counterexample in Remark~\ref{rem:nogo} of Sec.~\ref{append_sec:strong}, properties (v) and (vi) \emph{cannot} be extended to the entire class of ancilla-assisted linear-optical operations (those containing beam splitters coupling the system to classical ancillas): 
the measure $\NC$ does not vanish on classical states, and any operation class containing a channel that replaces the input by a fixed classical state necessarily violates monotonicity. 
The restriction in (v) and (vi) to the stochastic rotation--displacement class is therefore not a technical convenience but a structural necessity --- an observation that is fully consistent with, and indeed reinforces, the no-go theorem of the main text.

\section{Invariance of $\NC$ under rotation and displacement}
\label{append_sec:invariance}

\begin{definition} 
    The classicalization channel, $\Ccal$, maps any given state into a classical state such that, for an arbitrary density operator $\rho=\int\frac{\dd^2\alpha}{\pi}P(\alpha)\ket{\alpha}\!\bra{\alpha}$,
    \begin{align}
    	&\Ccal(\rho) = \Ccal\!\left[\int\frac{\dd^2\alpha}{\pi}P(\alpha)\ket{\alpha}\!\bra{\alpha}\right]=\int\frac{\dd^2\beta}{\pi}\Pcl(\beta)\ket{\beta}\!\bra{\beta} 
    	\nonumber 
    	\\
    	& = \int\frac{\dd^2\beta}{\pi}\left[\int\frac{\dd^2\alpha}{\pi}P(\alpha)f(\beta-\alpha)\right]\ket{\beta}\!\bra{\beta}
    	\nonumber \\
    	& = \int\frac{\dd^2\beta}{\pi}\int\frac{\dd^2\alpha}{\pi}\,P(\alpha)f(\beta)\ket{\beta+\alpha}\!\bra{\beta+\alpha}\nonumber\\
    	&=\int\frac{\dd^2\beta}{\pi}\int\frac{\dd^2\alpha}{\pi}\,P(\alpha)f(\beta)\,D(\beta)\ket{\alpha}\!\bra{\alpha}D^\dg(\beta)
    	\nonumber 
    	\\ 
    	&= \int\frac{\dd^2\beta}{\pi}\,f(\beta)\,D(\beta)\,\rho\,D^\dg(\beta)\,,
    	\label{append_eq:C}
    \end{align}
    where $\Pcl(\beta)=\int\frac{\dd^2\alpha}{\pi}P(\alpha)f(\beta-\alpha)\geq0$. Here, $f(\beta-\alpha)$ is a smoothing function applied on $P(\alpha)$ with $f(\beta)\geq0$.
\end{definition} 

Throughout this work the smoothing is taken to be \emph{isotropic}, $f(\beta)=f(|\beta|)$, as used for the Gaussian smoothing kernels employed in the main text.
This immediately makes $\Ccal$ commute not only with the averaged channel $\PhiL$ but with \emph{every individual} rotation--displacement unitary; this stronger, per-unitary statement is what is actually used in Secs.~\ref{append_sec:invariance}--\ref{append_sec:strong}, and we record it as a lemma.

\begin{lemma}[Per-unitary commutation]
\label{lem:commute}
    Let $V(\mu)=R(\theta)D(\lambda)$ and let $\Ccal$ be given by Eq.~\eqref{append_eq:C} with isotropic smoothing $f(\beta)=f(|\beta|)$. Then, for every state $\rho$,
    \begin{equation}  
        \Ccal\!\left[V(\mu)\,\rho\,V^\dg(\mu)\right] =V(\mu)\,\Ccal(\rho)\,V^\dg(\mu)\,.
    \label{append_eq:commute}
    \end{equation}
\end{lemma}

\begin{proof}
    Using $R^\dg(\theta)\,a\,R(\theta) = a\,\mathrm{e}^{\mathrm{i}\theta}$, one finds $R^\dg(\theta)D(\beta)R(\theta) = D(\beta\,\mathrm{e}^{-\mathrm{i}\theta})$, hence $D(\beta)R(\theta) = R(\theta)D(\beta\,\mathrm{e}^{-\mathrm{i}\theta})$. 
    Moreover, $D(\beta')D(\lambda) = \mathrm{e}^{\mathrm{i}\,\mathrm{Im}(\beta'\lambda^*)}D(\beta'+\lambda)$, and the scalar phases cancel between the left and right factors in $D(\beta')D(\lambda)\,\rho\,D^\dg(\lambda)D^\dg(\beta') = D(\beta'+\lambda)\,\rho\,D^\dg(\beta'+\lambda)$. 
    Therefore,
    \begin{align}
        &\Ccal\!\left[V\rho V^\dg\right] =\int\frac{\dd^2\beta}{\pi}f(\beta)\,D(\beta)R(\theta)D(\lambda)\,\rho\,D^\dg(\lambda)R^\dg(\theta)D^\dg(\beta)
        \nonumber
        \\
        &=\int\frac{\dd^2\beta}{\pi}f(\beta)\,R(\theta)\,D\!\left(\lambda+\beta\,\mathrm{e}^{-\mathrm{i}\theta}\right)\rho\,D^\dg\!\left(\lambda+\beta\,\mathrm{e}^{-\mathrm{i}\theta}\right)R^\dg(\theta) 
        \nonumber 
        \\
        &=\int\frac{\dd^2\beta'}{\pi}f\!\left(\beta'\mathrm{e}^{\mathrm{i}\theta}\right)R(\theta)\,D(\lambda+\beta')\,\rho\,D^\dg(\lambda+\beta')\,R^\dg(\theta)
        \nonumber
        \\
        &=V(\mu)\,\Ccal(\rho)\,V^\dg(\mu),
    \end{align}
    where the substitution $\beta'=\beta\,\mathrm{e}^{-\mathrm{i}\theta}$ preserves the measure and isotropy gives $f(\beta'\mathrm{e}^{\mathrm{i}\theta})=f(\beta')$.
\end{proof}

For any operator functional, whether linear or non-linear, $F$ in Hilbert space, $F(U\rho U^\dg)=UF(\rho)U^\dg$ is trivially true. This immediately leads to the covariance of the $p$-norm under any unitary evolution.
However, in the case of phase-space linear functionals ($F_L$), let us consider the form $F_L:= F^{G(\alpha)}_{\mathrm{L}}(\rho) = \Tr[\rho G(\alpha)]$, where $G(\alpha)$ is any operator parametrized by the phase-space variable~$\alpha$, such that $G(\alpha) = D(\alpha) G_0 D^\dagger(\alpha)$ and $G_0$ is rotationally-invariant, i.e., $\args{G_0,R(\theta)} = 0$.

\begin{lemma}
    Let $G_0$ be any bounded operator with $\args{G_0, R(\theta)} = 0 ~\forall ~\theta$ and $G(\alpha):= D(\alpha) G_0 D^\dagger(\alpha)$.
    Then, for any $U(z,\theta) = R(\theta)D(z)$,
    $U^\dagger(z,\theta) G(\alpha) U(z,\theta) = G\argp{e^{-i\theta}\alpha -z}$.
\end{lemma}

\begin{proof}
    It is straightforward to check that $R^\dagger(\theta) D(\alpha) R(\theta) = D(e^{-i\theta} \alpha)$.
    This immediately yields,
    \begin{align}
        &U^\dagger(z,\theta) G(\alpha) U(z,\theta) 
        \nonumber 
        \\
        &= D^\dagger(z) R^\dagger(\theta) D(\alpha) G_0 D^\dagger(\alpha) R(\theta) D(z)
        \nonumber 
        \\
        &= D^\dagger(z) D(e^{-i\theta}\alpha) G_0 D^\dagger(e^{-i\theta}\alpha) D(z)
        \nonumber 
        \\
        &= D(e^{-i\theta}\alpha - z) G_0 D^\dagger(e^{-i\theta}\alpha - z)
        = G(e^{-i\theta}\alpha - z).
    \end{align}
\end{proof}

This immediately leads to the result $U(z,\theta): F^{G(\alpha)}_{\mathrm{L}}(\rho) \to F^{G(e^{-i\theta}\alpha -z)}_{\mathrm{L}}(\rho)$.
All bona fide phase-space distributions can be recast in this form. For example, for a single-mode optical state, the Wigner and the Husimi functions can be written as
\begin{align}
    W(z) &=\Tr[\rho\,G_{\mathrm{w}}(z)] ; \quad G_{\mathrm{w}}(z)=2 \,D(z)\,\Pi\,D^\dg(z),
    \nonumber
    \\
    Q(z) &=\Tr[\rho\,G_{\mathrm{q}}(z)] ; \quad G_{\mathrm{q}}(z)= \ket{z}\!\bra{z},
    \label{append_eq:WQ}
\end{align}
where $D(z)=\mathrm{e}^{z\hat a^\dg-z^*\hat a}$ is the displacement operator, $\Pi$ is the parity operator and $\ket{z}$ is a coherent state.

As a consequence, the $p$-norm-based distance becomes
\begin{widetext}
	\begin{align}
		U(z,\theta): \left\|F^{G(\alpha)}_{\mathrm{L}}(\rho)-F^{G(\alpha)}_{\mathrm{L}}[\Ccal(\rho)]\right\|_p 
		& \to\;\left\|F^{G(\mathrm{e}^{-\mathrm{i}\theta}\alpha-z)}_{\mathrm{L}}(\rho)-F^{G(\mathrm{e}^{-\mathrm{i}\theta}\alpha-z)}_{\mathrm{L}}[\Ccal(\rho)]\right\|_p 
		\nonumber 
		\\
		&=\left[\int\dfrac{\dd^2\alpha}{\pi}\left|F^{A(\mathrm{e}^{-\mathrm{i}\theta}\alpha-z)}_{\mathrm{L}}(\rho)-F^{A(\mathrm{e}^{-\mathrm{i}\theta}\alpha-z)}_{\mathrm{L}}[\Ccal(\rho)]\right|^p\right]^{1/p}
		\nonumber 
		\\
		&= \args{ 
			\int\dfrac{\dd^2\alpha'}{\pi}\left|F^{G(\alpha')}_{\mathrm{L}}(\rho)-F^{G(\alpha')}_{\mathrm{L}}[\Ccal(\rho)]\right|^p
		}^{1/p}
		\nonumber 
		\\
		&= \args{ 
			\int\dfrac{\dd^2\alpha}{\pi}\left|F^{G(\alpha)}_{\mathrm{L}}(\rho)-F^{G(\alpha)}_{\mathrm{L}}[\Ccal(\rho)]\right|^p
		}^{1/p}
		= \left\|F^{G(\alpha)}_{\mathrm{L}}(\rho)-F^{G(\alpha)}_{\mathrm{L}}[\Ccal(\rho)]\right\|_p,
		\label{append_eq:volpreserve}
	\end{align}
\end{widetext}
since $\Ccal$ commutes with $U(z,\theta)$ by Lemma~\ref{lem:commute} and $\alpha\to\mathrm{e}^{-\mathrm{i}\theta}\alpha-z$ is a volume preserving symplectic transformation.
This completes the proof that $\NC$, which comprises linear functionals and the $p$-norm, remains invariant under rotations and displacements. 

\section{Strong monotonicity of $\NC$}
\label{append_sec:strong}

\begin{definition}[Free selective operations]\label{def:free}
A free selective operation is an instrument $\{\mathcal{E}_k\}$ of the form
\begin{equation}
    \mathcal{E}_k(\rho) = \int\dd\mu\,\Pcl(\mu)\,q_k(\mu)\,V(\mu) \rho\,V^\dg(\mu),\quad \argp{q_k(\mu)\geq 0},
    \label{append_eq:instrument}
\end{equation}
where $\sum_k q_k(\mu)= 1 ~\forall\mu$ and $V(\mu)=R(\theta)D(\lambda)$ are the rotation--displacement unitary operations, defined as Eq. \eqref{append_eq:commute}, and $q_k(\mu)$ is the conditional probability of announcing outcome $k$ given the realized parameter $\mu$---a classical post-processing (or coarse-graining of $\mu$). 
Equivalently, the Kraus operators are
\begin{equation}
    K_{k,\mu}=\sqrt{\Pcl(\mu)\,q_k(\mu)}\;V(\mu), \qquad \sum_k\int\dd\mu\,K_{k,\mu}^\dg K_{k,\mu}=\mathbb{I},
    \label{append_eq:kraus}
\end{equation}
\emph{i.e.}, each Kraus operator is proportional to a free unitary. The non-selective channel is recovered as $\sum_k\mathcal{E}_k=\PhiL$.
\end{definition}

Because every Kraus operator in Eq.~\eqref{append_eq:kraus} is proportional to a unitary, the outcome probabilities are state-independent,
\begin{align}
	p_k &=\Tr[\mathcal{E}_k(\rho)]=\int\dd\mu\,\Pcl(\mu)\,q_k(\mu) ,
	\nonumber 
	\\
	\rho_k &=\frac{\mathcal{E}_k(\rho)}{P_k}=\frac{1}{P_k}\int\dd\mu\,\Pcl(\mu)\,q_k(\mu)\,V(\mu)\,\rho\,V^\dg(\mu)\,.
	\label{append_eq:pk-rhok}
\end{align}

\subsection{Proof of strong monotonicity}
\label{append_subappend_sec:strong}

For notational compactness, define the linear map
\begin{align}
	G_\eta(\sigma) &:= F^{A(\eta)}_{\mathrm{L}}(\sigma)-F^{A(\eta)}_{\mathrm{L}}[\Ccal(\sigma)],
	\quad\text{so that}\quad
	\nonumber 
	\\
	\NC(\sigma) &=\|G_\eta(\sigma)\|_p=\left[\int\dd\eta\,|G_\eta(\sigma)|^p\right]^{1/p}.
	\label{append_eq:Gdef}
\end{align}

The linearity of $G$ in $\sigma$ follows from the linearity of $F^{A(\eta)}_{\mathrm{L}}$ and of the channel $\Ccal$. 
Moreover, by Lemma~\ref{lem:commute} and Eq.~\eqref{append_eq:volpreserve},
\begin{equation}
    \left\|G_\eta\!\left(V(\mu)\,\sigma\,V^\dg(\mu)\right)\right\|_p = \|G_\eta(\sigma)\|_p,
    \label{append_eq:Ginv}
\end{equation}
for every rotation--displacement unitary $V(\mu)$.

Now, using the positive homogeneity of the norm, $P_k\NC(\rho_k)=\|G(P_k\rho_k)\|_p = \| G[\mathcal{E}_k(\rho)] \|_p$, together with the linearity of $G$,
\begin{align}
    &\sum_k P_k\,\NC(\rho_k) 
    \nonumber 
    \\
    &= \sum_k\left\|\,\int\dd\mu \Pcl(\mu)\,q_k(\mu) G_\eta\args{ V(\mu)\,\rho\,V^\dg(\mu) } \right\|_p 
    \nonumber 
    \\
    &\leq\sum_k\int\dd\mu\,\Pcl(\mu)\,q_k(\mu)\left\| G_\eta\args{ V(\mu)\,\rho\,V^\dg(\mu) } \right\|_p
    \nonumber
    \\
    &=\sum_k\int\dd\mu\,\Pcl(\mu)\,q_k(\mu)\,\|G_\eta(\rho)\|_p
    \nonumber 
    \\
    &=\NC(\rho)\int\dd\mu\,\Pcl(\mu)\sum_k q_k(\mu)
    =\NC(\rho),
    \label{append_eq:strong}
\end{align}
where we have used the invariance in Eq.~\eqref{append_eq:Ginv} and the Minkowski inequality
\begin{equation}
    \normp{\int d\mu f(\mu) G(\mu)} \leq \int d\mu \abs{f(\mu)} \normp{G(\mu)}, \quad \argp{p\geq 1}
    \label{append_eq:Minkowski}
\end{equation} 
such that $f(\mu)$ is a measurable function. 
The last equality also uses $\sum_k q_k(\mu)=1$ and $\int\dd\mu\,\Pcl(\mu)=1$.

Note that the only inequality in Eq.~\eqref{append_eq:strong} is the Minkowski step. 
The inequality is saturated when, for each $k$, the functions $G(V(\mu)\rho V^\dg(\mu))$ coincide (up to a common phase pattern) on the support of $\Pcl(\mu)q_k(\mu)$, \emph{e.g.}, when each $q_k$ selects a single $\mu$ (a fine-grained readout of the classical noise), in which case $\rho_k=V(\mu_k)\rho V^\dg(\mu_k)$ and equality holds outcome by outcome. Genuine coarse-graining generically renders the inequality strict.

\subsection{No extension to ancilla-assisted linear optics}

One may ask whether Definition~\ref{def:free} can be enlarged to the physically natural class of \emph{ancilla-assisted} linear-optical operations: 
append a classical ancilla state $\sigma_{\mathrm{anc}}$ comprising $M$ modes \cite{Tan2017, Kuhn2018, Tan2020}, apply a passive-plus-displacement linear-optical unitary $U_{\vec\mu}$ (phase shifters, displacements \emph{and beam splitters}), and projectively measure the ancilla in an orthonormal basis $\{\ket{\Lambda_k}\}$,
\begin{equation}
    \Phi^{\Pi}_{\mathrm{L}}(\rho):=\sum_k\Tr_{\mathrm{anc}}\!\left[U_{\vec\mu}\,(\rho\otimes\sigma_{\mathrm{anc}})\,U^\dg_{\vec\mu}\,\Pi_k\right],
    \label{append_eq:ancilla-class}
\end{equation}
where $\Pi_k=\ket{\Lambda_k}\!\bra{\Lambda_k}\,$.

Every outcome of Eq.~\eqref{append_eq:ancilla-class} still maps classical states to classical states, so this class is classicality-preserving in the selective sense. 
Nevertheless, the answer is \emph{no}: $\NC$ is not monotone --- weakly or strongly --- under the class \eqref{append_eq:ancilla-class}. 
The obstruction is structural. 
We elaborate this situation through a counterexample.

\begin{remark}[Counterexample: the SWAP channel]
\label{rem:nogo}
\end{remark}
Let us consider a single mode ancilla at vacuum state. 
For a standard beam splitter, $U_\text{bs}(\pi/2)$, represents a swap-gate. 
In other words, $U_{\mathrm{bs}}(\pi/2)$ swaps the system state with the vacuum ancilla, i.e., 
\begin{equation}
    \Phi_{\mathrm{swap}}(\rho)=\ket{0}\!\bra{0}\qquad\text{for every input }\rho\,,
\end{equation}
and any subsequent ancilla measurement leaves the (product) system state untouched, so every conditional state equals the vacuum: $\rho_k=\ket{0}\!\bra{0}$ for all $k$. Hence
\begin{equation}
    \sum_k p_k\,\NC(\rho_k)=\NC(\ket{0}\!\bra{0})=:N_0>0\,,
\end{equation}
where $N_0>0$ because the fixed smoothing channel $\Ccal$ displaces even the vacuum away from itself, $\Ccal(\ket{0}\!\bra{0})\neq\ket{0}\!\bra{0}$.

Now exhibit a \emph{classical} input state whose measure lies strictly below this baseline. 
By the displacement covariance \eqref{append_eq:Ginv} and Lemma~\ref{lem:commute}, the coherent-state profile is a rigid translate of the vacuum profile,
\begin{equation}
    G(\ket{\alpha}\!\bra{\alpha})(\eta)=g(\eta-\alpha)\,,
    \quad g:=G(\ket{0}\!\bra{0})\,,
    \quad\|g\|_p=N_0\,.
\end{equation}

For the bona fide phase-space functionals of Eq.~\eqref{append_eq:WQ} with regular smoothing $f$, $g$ is a smooth, absolutely integrable function vanishing at phase-space infinity, so $g\in L^p$ for all $p\geq1$. Consider the balanced classical mixture
\begin{align}
	\rho_{\mathrm{mix}} &= \tfrac{1}{2}\left(\ket{\alpha}\!\bra{\alpha}+\ket{-\alpha}\!\bra{-\alpha}\right), 
	\nonumber 
	\\
	\Rightarrow G(\rho_{\mathrm{mix}})(\eta) &= \tfrac{1}{2}\left[g(\eta-\alpha)+g(\eta+\alpha)\right].
\end{align}

As $|\alpha|\to\infty$ the two translates have asymptotically disjoint supports, and dominated convergence gives
\begin{align}
    \args{ \NC(\rho_{\mathrm{mix}}) }^p &= \int\dd\eta\left|\tfrac{1}{2}\,g(\eta-\alpha) + \tfrac{1}{2}\,g(\eta+\alpha)\right|^p
    \nonumber 
    \\
    &\;\xrightarrow[|\alpha|\to\infty]{} \;2\cdot 2^{-p}\,\|g\|_p^p, 
    \nonumber
    \\
    \Rightarrow \NC(\rho_{\mathrm{mix}}) &\to 2^{(1-p)/p}\,N_0\,.
    \label{append_eq:mixture-limit}
\end{align}
For any $p>1$ one has $2^{(1-p)/p}<1$, so there exists $|\alpha|$ large enough that
\begin{equation}
    \sum_k p_k\,\NC(\rho_k)=N_0>\NC(\rho_{\mathrm{mix}})\,,
\end{equation}
violating strong monotonicity; since the outcome statistics are trivial, the same inequality $\NC[\Phi_{\mathrm{swap}}(\rho_{\mathrm{mix}})]>\NC(\rho_{\mathrm{mix}})$ violates weak monotonicity for the class \eqref{append_eq:ancilla-class} as well.
It may be noted that $p=1$ yields a trivial result as in that case $\left. \NC(\rho_{\mathrm{mix}}) \right|_{p=1} = N_0$.

\section{No-go theorem}
\label{append_sec:pneg}

\begin{definition}[$s$-ordered distribution and kernel]
	\label{def:s_ordered}
	Let $\mathcal{K}_s(\alpha) = D(\alpha)\,\mathcal{K}_s(0)\,D(\alpha)^\dagger$, $s\in[-1,1)$, where
	\begin{equation}
		\mathcal{K}_s(0) := \frac{2}{\pi(1-s)}\left(\frac{s+1}{s-1}\right)^{a^\dag a}=\frac{2}{\pi(1-s)}\,{:}e^{-\frac{2}{1-s}a^\dag a}{:}\,.
	\end{equation}
	This is the standard Cahill--Glauber $s$-ordered kernel; $s=0$ and $s=-1$ correspond to the familiar Wigner and Husimi
	kernels. 
    The following identity, standard in the phase-space formalism, see below
	\begin{align}
		W_s(\rho;\alpha)=&\,\mathrm{Tr}[\rho\,\mathcal{K}_s(\alpha)]=\frac{2}{\pi(1-s)}\mathrm{Tr}[\rho\,{:}e^{-\frac{2}{1-s}(a^\dag-\alpha^*) (a-\alpha)}{:}]\nonumber\\
		=&\,\dfrac{1}{\pi^2}\int d^2\lambda\,e^{-\frac{1-s}{2}|\lambda|^2}\mathrm{Tr}[\rho\,e^{\lambda(a^\dag-\alpha^*)}e^{-\lambda^*(a-\alpha)}]\nonumber\\
		=&\,\,\dfrac{1}{\pi^2}\int d^2\lambda\,e^{-\frac{1-s}{2}|\lambda|^2}\,e^{\lambda^*\alpha-\lambda\alpha^*}\,\mathrm{Tr}[\rho\,e^{\lambda a^\dag}e^{-\lambda^*a}]\nonumber\\
		=&\,\frac{1}{\pi^2}\int d^2\lambda\;\chi_s(\lambda;\rho)\,e^{\lambda^*\alpha-\lambda\alpha^*}\,,
	\end{align}
	where we have used the identity
	\begin{equation}
		\frac{1}{\pi}\int d^2\lambda\,e^{-a|\lambda|^2+b_1\alpha+b_2\alpha^*}=\dfrac{1}{a}\,e^{\frac{b_1b_2}{a}}
	\end{equation}
	to write $W_s(\rho;\alpha)$ as a two-dimensional Fourier transform of the characteristic function $\chi_s(\lambda;\rho)=\mathrm{Tr}[\rho\,D(\lambda)]\,e^{s|\lambda|^2/2}$.
\end{definition}

\begin{definition}[Gaussian noise-additive classicalization channel]
	\label{def:gaussian_classicalization}
	A Gaussian noise-additive classicalization channel $\mathcal C_g$ is a subset of all classicalization channels $\mathcal{C}$ \eqref{append_eq:C}, defined as the
	quantum-limited attenuator--amplifier composite channel, i.e., $\mathcal C_g=A_2\circ E_{1/2}$, where
    \begin{align}
    \mathcal{E}_\lambda \argp{\rho} &= \rm{Tr}_\text{anc} \args{ \mathsf{B}(\lambda) \argp{ \rho\otimes \ket{0_\text{anc}}\bra{0_\text{anc}} } \mathsf{B}^{\dagger}(\lambda) } ~~\rm{and}
    \nonumber 
    \\
    \mathcal{A}_k \argp{\rho} &= \rm{Tr}_\text{anc} \args{ \mathsf{S}(\mathit{k}) \argp{ \rho\otimes \ket{0_\text{anc}}\bra{0_\text{anc}} } \mathsf{S}^{\dagger}(\mathit{k}) },
    \label{eq:defn_attenuate_amplify}
\end{align}
such that $\mathsf{B}(\lambda) = e^{\cos^{-1}\sqrt{\lambda}(a^\dagger b - a b^\dagger)}$ is a passive beam-splitter with transmittivity $\lambda$, and $\mathsf{S}(k) = e^{\cosh^{-1}\sqrt{k}(a^\dagger b^\dagger - a b)}$ is a two-mode correlated squeezing operator. 
Here $\ket{0_\text{anc}}$ denotes the vacuum state in the ancilla mode.

For $\mathcal{C}_g$, $f(\beta)=e^{-|\beta|^2}$ such that $P_{\mathcal{C}_g(\rho)}(\alpha) = Q_\rho(\alpha)$, the Husimi function of $\rho$ --- equivalently, $P$ convolved with a Gaussian of variance~$1$ \cite{Linowski2024}.
\end{definition}

\begin{lemma}
	\label{lem:Cgtherm}
	For the thermal state $\rho_{th}(\bar n)$, $\chi(\lambda;\rho_{th}(\bar n))=e^{-(\bar n+1/2)|\lambda|^2}$, $s\in [-1,1)$, so
	\begin{equation}
		W_s\args{ \rho_{th}(\bar n);\alpha } = \frac{1}{\pi\tau_s(\bar n)}\,e^{-|\alpha|^2/\tau_s(\bar n)}\,,
        \label{append_eq:sdist_therm}
	\end{equation}
	with $\tau_s(\bar n):=\frac{2\bar n+1-s}{2}\,$ is a genuine convergent Gaussian for every $s<2\bar n+1$ (in particular every $s\in[-1,1)$), $\exists ~0< n^{*} < \infty$ such that 
    \begin{equation}
        \NCs\args{\rho_\text{th}(\bar{n})} < \frac{\bar{n} - 1}{\bar{n} + 1} N_0 \quad;\quad \forall ~\bar{n}>n^{*},
        \label{append_eq:sdistrhoth_ub}
    \end{equation}
    where $N_0 = \NCs(\ket 0)$. 
\end{lemma}

\begin{proof}
    By Definition \ref{def:gaussian_classicalization}, the Gaussian noise-additive channel, $\mathcal{C}_g$ that convolves $P$ with a Gaussian of variance~$1$, increases the variance by one unit.
    As a consequence, under the operation of $\mathcal{C}_g$, the s-ordered distribution of a thermal state \eqref{append_eq:sdist_therm} becomes another thermal state where the average photon-number is increased by unity, i.e.,
    \begin{equation}
        \mathcal{C}_g: \rho_{th}(\bar n) = \mathcal{C}_g \args{\rho_{th}(\bar{n})} = \rho_{th}(\bar{n} + 1)
    \end{equation}

    We now prove the lemma in two cases.\\
    
    {\bf Case 1: $p=1$}\\

    For $\tau>0$, writing $W_\tau(\alpha):=\frac1{\pi\tau}e^{-|\alpha|^2/\tau}$, the two Gaussian functions $W_\tau$ and $W_{\tau+1}$ are positive, normalized, and differ in sign exactly once radially, whose difference has total vanishing integral; integrating exactly gives
	\begin{equation}
		\big\|W_\tau - W_{\tau+1}\big\|_1 = \frac{2}{\tau+1}\left(\frac{\tau}{\tau+1}\right)^{\!\tau}\,.
		\label{append_eq:norm1_gaussdiff}
	\end{equation}

    This immediately leads to the results
    \begin{subequations}
        \begin{equation}
            \NCwo\big(\rho_{th}(\bar n)\big) = \frac{2}{\tau_s(\bar n)+1}\left(\frac{\tau_s(\bar n)}{\tau_s(\bar n)+1}\right)^{\!\tau_s(\bar n)}\,,
            \label{eq:NCthermal}
        \end{equation}
        \begin{equation}
            N_0^1(s) := \NCwo(|0\rangle) = \frac{2}{\tau_s(0)+1}\left(\frac{\tau_s(0)}{\tau_s(0)+1}\right)^{\!\tau_s(0)}\,,
            \label{eq:NCvacuum}
        \end{equation}
    \end{subequations}
    with $\tau_s(0)=\frac{1-s}{2}$ such that $N_0(s)$ is strictly positive $\forall ~s\in[-1,1]$.
    Moreover, $\lim_{\bar{n}\to\infty} \NCs(\rho_{th}(\bar n))\to0$ while $\lim_{\bar{n}\to\infty} \frac{\bar n-1}{\bar n+1}N_0(s)\to N_0(s)>0$. 
    Hence $\exists ~\bar n^*(s)<\infty$ such that
	\begin{equation}
		\NCwo\big(\rho_{th}(\bar n)\big) < \frac{\bar n-1}{\bar n+1}N_0^1(s) \qquad \text{for all } \bar n>\bar n^*(s)\,.
		\label{append_eq:nogocond_therm_p1}
	\end{equation}

    {\bf Case 2: $p\geq 2$}\\

    By writing $|\alpha|^2 = x$, after angular integration we have $\frac{d^2\alpha}{\pi} = dx$. 
    As a consequence, the $\mathcal{L}^p$ distance between the Gaussian functions $W_\tau$ and $W_{\tau+1}$ is given as
    \begin{align}
        &N_p(\tau) :=\NCs\argp{\rho_\text{th}(\bar{n})} 
        \nonumber 
        \\ &= \args{ \int_0^{\infty} dx \abs{\frac{1}{\tau} e^{-x/\tau} - \frac{1}{\tau+1} e^{-x/(\tau+1)}}^p
        }^{1/p}
        \nonumber
        \\
        &= \tau^{1/p - 1} \args{ \int_0^{\infty} dy \abs{ e^{-y} - \frac{\tau}{\tau+1} e^{-\frac{\tau}{\tau+1}y} }^p
        }^{1/p},
        \quad 
        \argp{x=\tau y}.
        \label{append_eq:nptherm_ini}
    \end{align}

    Now, in the large $\bar{n}$ limit
    \begin{equation}
        \lim_{\tau\to\infty} \frac{\tau}{\tau + 1} = 1 - \frac{1}{\tau} + \mathcal{O}\argp{\tau^{-2}}
    \end{equation}
    leading to 
    \begin{equation}
        e^{-y} - \frac{\tau}{\tau+1} e^{-\frac{\tau}{\tau+1}y} \to \frac{1}{\tau}(1-y)e^{-y} + \mathcal{O}\argp{\tau^{-2}}.
    \end{equation}

    Substituting Eq. \eqref{append_eq:nptherm_ini} we obtain
    \begin{align}
        N_p(\tau) &\sim \tau^{1/p-2} \args{ \int_0^{\infty} dy \abs{ 1- y }^p e^{-py}
        }^{1/p}
        \nonumber 
        \\
        \Rightarrow \NCs\argp{\rho_\text{th}(\bar{n})} &= \mathcal{O} \argp{\tau_s(\bar{n})^{1/p-2}}.
        \label{append_eq:nptherm}
    \end{align}

    Evidently, $\forall ~p\geq 2$, $\lim_{\bar{n}\to\infty} \NCs\argp{\rho_\text{th}(\bar{n})} \to 0$ (as $1/p-2<0$), while $\lim_{\bar{n}\to\infty} \frac{\bar{n}-1}{\bar{n}+1}N_0^p(s) \to N_0^p(s)$, where $N_0^p(s) = \NCs(\ket 0)$.
    This, in turn, indicates that $\forall ~p\geq 2$, $\exists ~n^{*}<\infty$ such that
    \begin{equation}
		\NCs\big(\rho_{th}(\bar n)\big) < \frac{\bar n-1}{\bar n+1}N_0^p(s) \qquad \text{for all } \bar n>\bar n^*(s)\,.
		\label{append_eq:nogocond_therm_pg1}
	\end{equation}

    Thus, we have a precise result that $\forall ~s \in [-1,1)$, $\forall ~1\leq p< \infty$, $\exists ~\bar{n}^{*}(p,s)$ such that
    \begin{equation}
        \NCs\argp{\rho_\text{th}(\bar{n})} < \frac{\bar{n}-1}{\bar{n}+1} N_0^{p}(s), \quad \forall ~\bar{n}>n^{*},
        \label{append_eq:nogocond_therm}
    \end{equation}
    where $N_0^{p}(s) = \NCs(\ket 0)$.
    This concludes the proof.
\end{proof}

\begin{theorem}[No-go theorem]
\label{thm:nogo}
For the $s$-ordered quasidistribution $W_s(\rho;\alpha)=\mathrm{Tr}[\rho\,\mathcal{K}_s(\alpha)]$ and the $L^p$ measure $\NCs$ of Definition~\ref{def:gaussian_classicalization}, the certification criterion above cannot hold
universally: there exists a quantum state $\rho_{\rm neg}^{\rm test}$ that is genuinely $P$-negative yet satisfies
$M(\rho_{\rm neg}^{\rm test})<0$, where $M(\rho):=\NCs(\rho)-\NCs(|0\rangle)$.
\end{theorem}

\begin{proof}
	Let us consider a vacuum subtracted thermal state as
	\begin{equation}
		\rho_{\rm neg}^{\rm test} := \left(1+\frac1{\bar n}\right)\rho_{th}(\bar n) - \frac1{\bar n}|0\rangle\langle0|\,
	\end{equation}
	which represents a normalized physical state for every $\bar n>0$, with
	\begin{equation}
		P_{\rho_{\rm neg}^{\rm test}}(\alpha) = \frac{1+\bar n}{\bar n^2}\,e^{-|\alpha|^2/\bar n} - \frac{\pi}{\bar n}\,\delta^{(2)}(\alpha)\,,
	\end{equation}
	carrying a negative delta spike at the origin for every finite $\bar n>0$: $\rho_{\rm neg}^{\rm test}$ is genuinely
	$P$-negative, for every $\bar n>0$, independent of $s$.

	By writing $h^s_\rho(\alpha):=W_s(\rho;\alpha)-W_s(\mathcal C_g(\rho);\alpha)$, linearity of $W_s$ and of $\mathcal C_g$
	(Definitions~\ref{def:s_ordered} and~\ref{def:gaussian_classicalization}) gives
	\begin{equation}
		h^s_{{\rho_{\rm neg}^{\rm test}}} = \argp{ 1+\frac1{\bar n} } h^s_{\rho_{th}(\bar n)} - \frac1{\bar n}h^s_{\ket 0}\,
	\end{equation}
	leading to the result
    \begin{align}
        \NCs \argp{ \rho_\text{neg}^\text{test} } &= \normp{h^s_{\rho_\text{neg}^\text{test}}}
        \nonumber 
        \\
        &= \normp{ \argp{ \frac{1}{\bar{n} } } h^s_{\rho_\text{th}(\bar{n})} - \frac{1}{\bar{n}} h^s_{\ket 0} }
        \nonumber 
        \\
        &\leq \argp{ \frac{1}{\bar{n} } }  \normp{ h^s_{\rho_\text{th}(\bar{n})} } + \frac{1}{\bar{n}} \normp{ h^s_{\ket 0} }
        \nonumber 
        \\
        &= \argp{ \frac{1}{\bar{n} } } \NCs\argp{ \rho_\text{th}(\bar{n}) } + \frac{1}{\bar{n}} \NCs(\ket 0),
    \end{align}
    where we have used the triangle inequality $\normp{A-B} \leq \normp{A} + \normp{B}$.
    Evidently by Lemma \ref{lem:Cgtherm}, $\forall ~1\leq p<\infty$, $\forall ~\bar{n}>n^{*}$ 
    \begin{align}
    	\NC(\rho_{\rm neg}^{\rm test}) &<  \frac{1}{\bar{n}} \frac{\bar{n}-1}{\bar{n}+1} N_0^p(s) + \frac{1}{\bar{n}} N_0^p(s)
    	\nonumber 
    	\\
    	&= N_0^p(s) = \NCs(\ket 0),
    	\label{append_eq:nogo_sdist}
    \end{align}
	i.e., for some finite $\bar{n}$, the state $\rho_{\rm neg}^{\rm test}$ is $P$-negative but satisfies $\NCs(\rho_{\rm neg}^{\rm test}) < \NCs(|0\rangle)$, leading to $M(\rho_{\rm neg}^{\rm test})<0$.
\end{proof}

\subsection{An open extension of the no-go theorem---yet to be proved}

Consider a classical state, $\rho_\text{cl} = \int\frac{d^2\alpha}{\pi} P_\text{cl}(\alpha) \ket\alpha\bra\alpha$ ($P_\text{cl}(\alpha) \geq 0$) such that $\NC(\rho_\text{cl}) < \NC(\ket 0)$ for any $F_L:=\rm{Tr}\args{\rho G(\alpha)}$ where the generator $G(\alpha)$ is a displaced rotationally-invariant operator and $\mathcal{C}$ is any \emph{classicalization} channel. 
One can construct a normalised vacuum-subtracted classical state as
\begin{equation}
    \rho_\text{vsc} = \frac{1}{1-p_0} \rho_\text{cl} - \frac{p_0}{1-p_0} \ket 0\bra 0
    \label{append_eq:rhovsc}
\end{equation}
for which the corresponding $P$-function, given as
\begin{equation}
    P_\text{vsc}(\alpha) = \frac{1}{1-p_0} P_\text{cl}(\alpha) - \pi \frac{p_0}{1-p_0} \delta^2(\alpha),
    \label{append_eq:Pvsc}
\end{equation}
fails to maintain strict pointwise positivity, implying that $\rho_\text{vsc}$ is a nonclassical state, $p_0 = \text{Tr} \args{\rho_\text{cl} \ket 0\bra 0}$.
Following a calculation similar to that for $\NCs(\rho_\text{neg}^\text{test})$, using the definition in Eq. \eqref{append_eq:Gdef} and the triangle inequality, one can show that 
\begin{align}
    \NC(\rho_\text{vsc}) 
    \label{append_eq:nogo}
\end{align}
whenever
\begin{equation}
    \NC(\rho_\text{cl}) < (1-2p_0) \NC(\ket 0).
    \label{append_eq:nogo_cond}
\end{equation}

The existence of the no-go criterion in Eq. \eqref{append_eq:nogo_cond} can be analytically established for a subclass of $F_L$---all s-ordered distributions where $s\in[-1,1)$ and a subclass of classicalization channels---Gaussian unit-noise additive channel as done in Lemma \ref{lem:Cgtherm}.
However, it remains an open question whether Eq. \eqref{append_eq:nogo_cond} can be shown in general.

\end{document}